\documentclass[dvips]{arxstspdf}
\usepackage{graphicx}
\usepackage{flushend}
\usepackage{stfloats}

\volume{26}
\issue{1}
\pubyear{2011}
\firstpage{150}
\lastpage{159}
\doi{10.1214/09-STS300}

\begin{document}
\begin{frontmatter}

\title{A Random Walk with Drift: Interview with~Peter~J.~Bickel}
\runtitle{Interview with Peter J. Bickel}

\begin{aug}
\author{\fnms{Ya'acov} \snm{Ritov}\corref{}\ead[label=e1]{yaacov@mscc.huji.ac.il}}
\runauthor{Y. Ritov}

\address{Ya'acov Ritov is Professor, Department of Statistics, The Hebrew University of Jerusalem, Mount Scopus, Jerusalem,
91905, Israel
\printead{e1}.}
\end{aug}



\end{frontmatter}

I met Peter J. Bickel for the first time in 1981. He came to Jerusalem
for a year; I had just started working on my Ph.D. studies. Yossi Yahav,
who was my advisor at this time, busy as the Dean of Social Sciences,
brought us together. Peter became my chief thesis advisor. A year and a
half later I came to Berkeley as a post-doc. Since then we have
continued to work together. Peter was first my advisor, then a teacher,
and now he is also a friend. It is appropriate that this interview took
place in two cities. We spoke together first in Jerusalem, at Mishkenot
Shaananim and the Center for Research of Rationality, and then at the
University of California at Berkeley. These conversations were not
formal interviews, but just questions that prompted Peter to tell his
story.

The interview is the intellectual story of a post-war Berkeley
statistician who certainly is one of the leaders of the third generation
of mathematical statisticians, a~generation which is still fruitful
today.

The conversation was soft spoken, a stream of memories, ordered more by
association than by chro\-nology. In fact, I led Peter to tell his story
in a reverse direction, starting from the pure science and ending with
the personal background. So, please sit back, and imagine you are part
of the chat.

\textit{Peter, if you try to summarize the many stages of your career,
how do you characterize the different periods?}

A random walk with drift. Shall I start with the very beginning?

\textit{No, for now can you tell us about your academic career?}

I would say that the first period was almost exclusively theoretical
work, but actually, I think, almost from the beginning, driven as much
by people as by a focus on the subject.

I did my thesis with Erich Lehmann. From the thesis, I published two
papers on multivariate analogues of Hotelling's $T^{2}$ (Bickel, \citeyear{1964};
Bickel, \citeyear{1965a}), really not knowing much about multivariate analysis at
all, learning asymptotics as I went along. After the thesis, partly
talking with Peter Huber, and partly talking with Erich Lehmann, I did
some interesting things on questions of robust estimation. I had a paper
on trimmed means and how they compare to the mean and median (Bickel,
\citeyear{1965b}); again, the results were in the spirit of Hodges and Lehmann.

\begin{figure}

\includegraphics{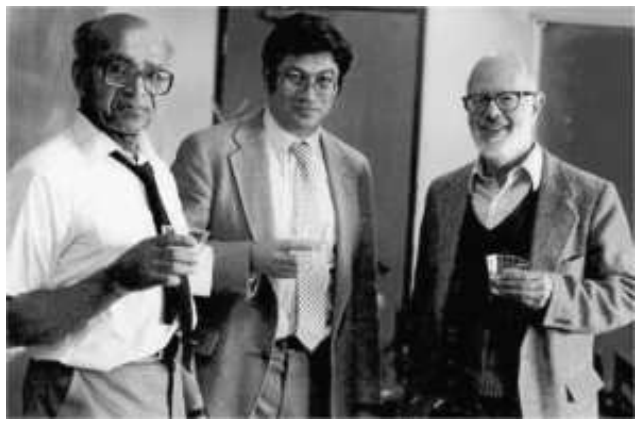}

  \caption{David Blackwell, Peter Bickel and Erich Lehmann at a party
celebrating Peter's election the National Academy of Science.}
\end{figure}

The next stage happened by a curious accident due to Govindarajulu. He
asked me if I had ever thought about investigating linear combinations
of order statistics. I said, no, but I had some ideas, having learned
about weak convergence of stochastic processes. Initially, he said he
wanted to work with me, but, at the same time, he was talking with
people in Stanford; and then he carried the problem to Le Cam. The
result was, finally, that the problem was attacked with three different
approaches. One was the approach of the Stanford group, growing from the
work Herman Chernoff did on rank statistics, mine, using weak
convergence of the quantile process (Bickel, \citeyear{1967Bickel}), and Le Cam's, which
used the H\'{a}jek projection technique. We all got
results.

Within this work, I was very pleased about the result that the
covariance of two order statistics is non-negative, which Richard Savage
claimed was a~long unsolved problem. Then it turned out that it was an
inequality in Hardy, Littlewood and Polya. This work also led to work in
multivariate goodness of fit tests and other problems to which I applied
the new notions of weak convergence of processes.

I went through my Ph.D. studies very quickly, which left me unfamiliar
with many parts of statistics. I tried to fill the gaps later on. After
my thesis Yossi Yahav and I started talking about his notion of
asymptotic pointwise sequential analysis purely from a theoretical point
of view. He had solved some special cases, and I realized that there was
a general pattern we could use. We have two or three papers in that
direction (Bickel and Yahav, \citeyear{1967BickelYahav}; Bickel and Yahav,
\citeyear{1968}; Bickel and
Yahav, \citeyear{1969a}; Bickel and Yahav, \citeyear{1969b}).

Erich Lehmann characterizes people as problem solvers, like Joe Hodges,
and system builders, like Erich himself. I fall somewhere in between,
but I am primarily a problem solver.

\begin{figure}[b]

\includegraphics{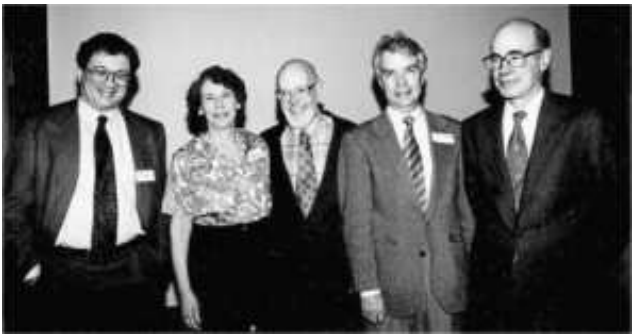}

  \caption{From left to right: Peter Bickel, Juliet Shaffer, Erich
Lehmann, Kjell Doksum and David Freedman, celebrating the 65th birthday
of Erich Lehmann.}
\end{figure}

An area that I started to work on in the seventies with van Zwet and
G\"{o}tze was second order asymptotics (Bickel, G\"{o}tze and van Zwet,
\citeyear{1985}, \citeyear{1986}). It was prompted by Hodges and Lehmann's paper on
deficiency. I got an idea on how to prove things for one sample rank
tests. Van Zwet, independently, got further, but was stumped by two
samples tests. We talked and, using a method of Renyi, we got a complete
answer for rank test statistics and permutation test statistics (Albers,
Bickel and van Zwet, \citeyear{1976}). Later, I was asked to give what is now known
as an IMS Medallion Lecture, for which I had to give a~topic. I proposed
Edgeworth expansions and nonparametric statistics, some of which I knew
how to do, at least formally. But then I had to have a~real example, so
I chose $U$-statistics. Something like a~month before the lecture I~realized
that there was a substantial difficulty with my arguments.
Luckily, just in time, I found an idea which worked. Not quite the right
idea, but it did give the first Berry--Esseen bound for $U$-statistics
(Bickel, \citeyear{1974}). Subsequently, Jon Weirman in Seattle got it right.

I took part in the Princeton Robustness year in 1971 (Andrews et al.,
\citeyear{1972}). Unfortunately, I didn't understand Tukey most of the time. He had
his own language that one had to follow. But there were other
interesting people I could talk with easily. Peter Huber was there,
David Andrews, and Frank Hampel. One of the things that came up was the
issue of adaptation. Peter Huber and I agreed that it was symmetry that
was doing the trick. Then I decided to think more about this question.
But we'll return to that in a moment.

Then, surprisingly, I moved into something genuinely applied.

\begin{figure}

\includegraphics{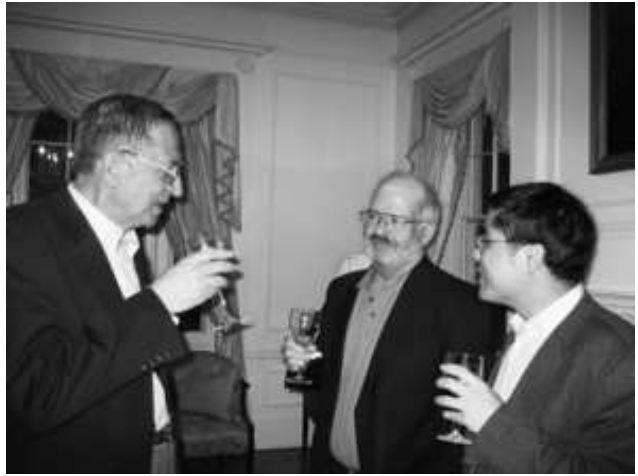}

  \caption{Willem van Zwet, Jon Wellner and Jianqing Fan. Bickelfest,
Princeton 2005.}
\end{figure}

Somehow, through Joe Hodges, I got interested in finding out more about
the university, so I joined something called the Graduate Council.
Eugene Hammel, the Associate Dean of Graduate Studies, presi\-ded over the
Council. One day, Hammel told me he had a very strange problem: he had
analyzed data on graduate admissions, because he was worried that the
government would cut funding, on the grounds that Berkeley was biased.

Indeed, he found strong evidence of gender bias. So he looked to find
the departments or units where the bias was, since decisions were made
on the department or unit level; he couldn't find them. I told him that
there is no contradiction and I gave him an example of what I later
learned was the Simpson paradox. Eventually, we made several tests for
conditional independence by units, one of which we later found out was
equivalent to the Mantel-Haenszel test. It was an enjoyable paper; it
appeared in \textit{Science} (Bickel, Hammel and O'Connell, \citeyear{1975}).

I had some other excursions into applications. I was recommended by
Betty Scott to the National Research Council and served on two
committees\break which studied two insurance problems. The first problem was
how to implement a mud slide insurance program. The government already
ran a flood insurance program in which it subsidized insurance companies
to give insurance to flood areas, provided the communities would agree
that people would not build on the flood plains. When Southern
California had a period of extreme rain, there were mud slides all over.
Consequently, the representatives of the districts with mud slides got a
law passed in Congress requiring the government to construct a~mud slide
insurance program. But nobody knew how to do~that.

\begin{figure}

\includegraphics{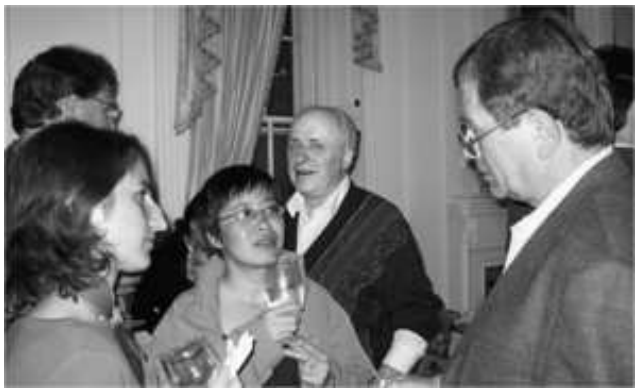}

  \caption{Katerina Kechris, Haiyan Huang, Friedrich Goetze and Peter
Bickel. Bickelfest, Princeton 2000.}
\end{figure}

So they convened the mud slide panel. It turned out that the main
problem was the nature of the data. They had very extensive aerial
photographs of the extent of mud slides, but nobody knew whether they
had happened last year or a thousand years ago, so there was no way to
set premiums. Basically, the panel knew what had to be done. They
proposed that teams of engineers be collected to look at the candidate
areas. The engineers would scratch their heads, and they would come up
with an insurance rate, given the information available. I couldn't see
what else could be done. However, I suggested, ``Why don't you have
different groups of engineers rate the same area and see if you can test
for consistency?'' Nobody else on the panel agreed.

The funny thing is that a year or two later I was put on another panel
dealing with the same issue. This time it was about flood insurance. The
problem was that the Federal Insurance Administration contracted out to
different government agencies to assess flood risk. The US Geological
Survey and the Army Corps of Engineers were assessing adjacent areas.
The border was in the middle of a flood plain, but they came up with
different rates for the two halves of the flood plain! Anyway that was
fun. I enjoyed it. But I never did any serious data analysis, in part
because I didn't trust myself to be sufficiently observant.

\begin{figure}

\includegraphics{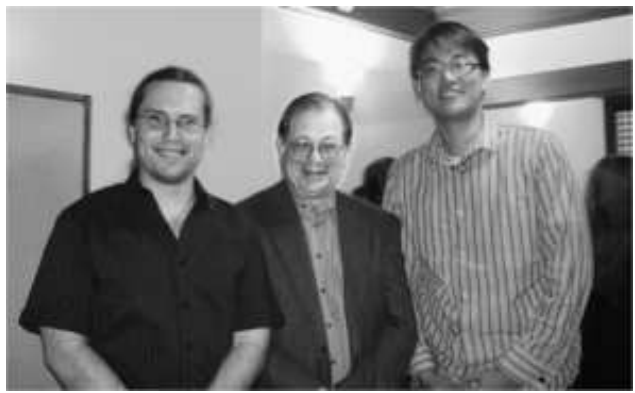}

  \caption{Peter Bickel and two recent students, Ben Brown (left) and
Choongsoon Bae.}
\end{figure}

Another line of research that had an impact on my later interests was a
paper on the maximum deviation of kernel density estimates that I worked
on with Murray Rosenblatt (Bickel and Rosenblatt, \citeyear{1973}). Murray came to
my office back in the 70s and asked whether empirical process theory
with weak convergence would work on extrema. I started to think about
this and realized that it couldn't work that way, because the limit is
white noise. Eventually I saw a way that you could attack the problem
with Skorohod embedding. While working with Rosenblatt and van Zwet, I
read an old paper of Hodges and Lehmann where they looked at minimaxity
subject to restrictions, where they did some explicit calculations for
the binomial. At some point I learned an identity that Larry Brown
pointed to: you can relate the Bayes risk in the Gaussian shift model to
the Fisher information of the marginal distribution. This led to papers
on semiparametric robustness (Bickel, \citeyear{1984}) and on estimation of a
normal mean (Bickel, \citeyear{1983}), which had a surprising follow-up in the work
of Donoho and Johnstone.

At about the same time, I looked at the question of adaptation again. A
paper on that subject (Bickel, \citeyear{1982}), as well as the preceding work on
asymptotic restricted minimax, was developed during a Miller
professorship, and that work was given in my Wald lectures. Then I came
to Israel on sabbatical and I had the good fortune of having you as a
student working on Bayesian robustness.

The next stage started when I gave lectures at Johns Hopkins on
semiparametric models, and I began to put things in context and realized
the connections with Jon Wellner's and Pfanzagl's work and robustness
and so on. Then you, Jon, Chris Klaassen and I started working on our
joint book (Bickel et al., \citeyear{19931998}) and, in the process, solved,
separately or jointly, the problems of censored regression and
errors-in-variables (Bickel and Ritov, \citeyear{1987}). It was great fun working
on the book.

A significant excursion was some work with Leo Breiman. Leo was visiting
Berkeley, considering whe\-ther to become a faculty member. He spoke about
a multivariate goodness of fit test he devised using nearest neighbors
in high dimension. Its asymptotic limiting behavior was harder to
understand than either of us thought initially, but eventually we worked
it out (Bickel and Breiman, \citeyear{1983BickelBreiman}). The work appeared in the
\textit{Annals of Probability} because the \textit{Annals of Statistics}
was then run by David Hinkley, whose views of what constituted
statistics were quite different from mine.

\begin{figure}[b]

\includegraphics{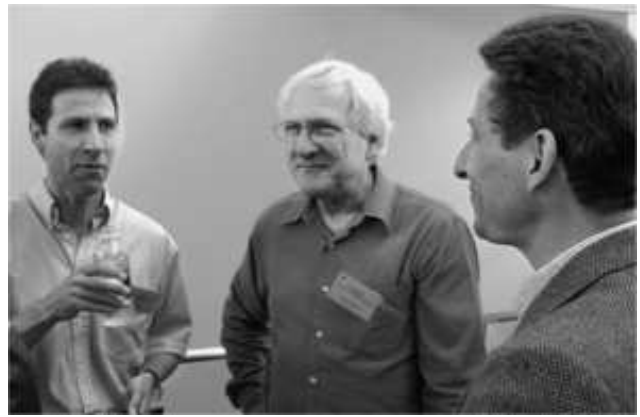}

  \caption{Peter Buehlmann, John Rice and Niklaus Hengartner. Bickelfest,
Princeton, 2005.}
\end{figure}

Another excursion which was important was to the theory of the
bootstrap, on which I early on worked with Freedman, and later in the
90s with G\"{o}tze and van Zwet (Bickel, G\"{o}tze and van Zwet, \citeyear{1995Bickel}).
Brad Efron introduced the bootstrap, following his profound insight into
the impact of computing on statistics. It took me a while to realize
that the bootstrap could be viewed as Monte Carlo implementation of
nonparametric maximum likelihood. So these various things intertwined. I~eventually became interested in the interplay between high dimensional
data and computing and the tradeoff between computing and efficiency.
From the 90s to the present, you and I moved from semiparametrics to
nonparametrics, for example, nonparametric testing, the LASSO and that
kind of thing. I think it is fair to say that this work was promoted in
part by our participation in an unclassified National Security
Administration program and in part by conversations with Leo Breiman. As
you know, Leo and I got quite close. I~learned a lot from him and I
became more sharply aware that high dimensional data and computing had
led to a paradigm change in statistics.

During our collaboration, I have tried to keep up with you. Working with
graduate students---especial\-ly you---is a key part of my life. Ideas
come to me when I talk. I have had lots of students, many very good, a
few outstanding; all of them were important to me. Just as important
have been senior collaborators, including a number of former students
and colleagues at Berkeley, Chicago, Stanford, Seattle, Michigan,
Harvard, Zurich, Leiden, Bielefeld and Israel.

\begin{figure}[b]

\includegraphics{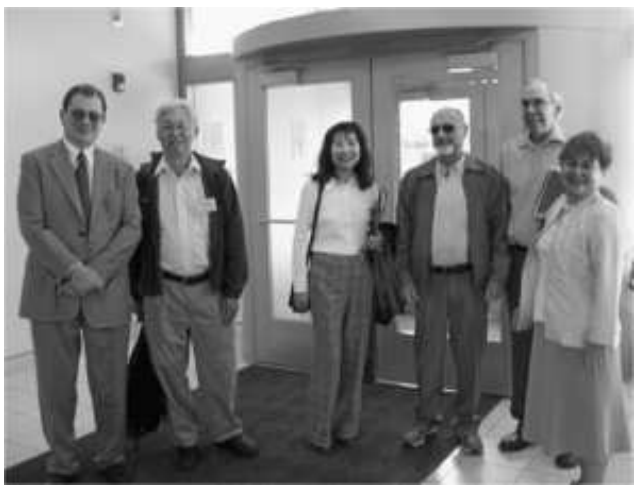}

  \caption{Peter Bickel, Quang Pham, Kang James, Yossi Yahav, Berry
James and Nancy Bickel. Bickelfest, Princeton 2005.}
\end{figure}

As I became older, I finally became bolder in starting to think
seriously about the interaction between theory and applications, a
direction initiated in part by working with my student and, later,
collaborator, Liza Levina. Nancy, my wife, thinks, and I think she is
right, that getting the MacArthur Fellowship made a change. I never was
sure of myself and the MacArthur helped. I became more self-confident.
When my student, Niklaus Hengartner, was working on a specific applied
problem, we realized that the theoretical semiparametric ideas really
helped (Hengartner et al., \citeyear{1995Hengartner}). Then, the work
with John Rice, you, and the Engineering and Computer Science people on
transportation problems played an important role. I think my
collaboration with John Rice has been very fruitful. He is a wonderful
data analyst, has very interesting ideas on the questions that can be
asked and is also very knowledgeable about different techniques. I
really like the recent work with John and Nicolai Meinshausen, which is
coming out in the \textit{Annals of Applied Statistics} (Meinshausen,
Bickel and Rice, \citeyear{2009}). As far as I know, it is the first paper which
makes it absolutely clear that a main issue is the tradeoff between the
efficiency of a procedure and the amount of computer time required to
implement it successfully.

\begin{figure}

\includegraphics{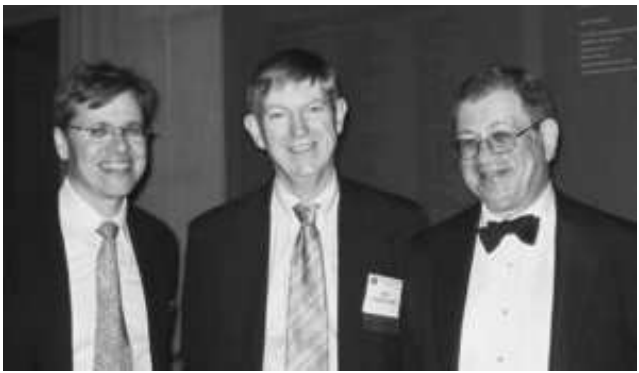}

  \caption{David Donoho, Iain Johnstone and Peter Bickel, celebrating
Iain's election to the National Academy.}
\end{figure}

Then there is biology. I was always interested in biology. I met this
wonderful guy, Alexander Glazer, who was then the chair of Molecular
Biology in Ber\-keley. At some point, because I was thinking about
exploring biology, we talked after a lecture. He said he was unhappy
about critiques by phylogeneticists of recent work on some proteins he
had long studied. When he gave a talk at Stanford, these critics claimed
that, using statistical methods, they had obtained a phylogenetic tree
that contradicted his views. I had some doubts about statistical methods
in this context, said so, and we started to talk. He was just retiring
and closing his lab, becoming a high level administrator, but he wanted
to keep his hand in.

I had a very good student, Katerina Kechris, who was just starting. I
got her involved in his program. I told her it was risky, but we had a
chance to work with a real biologist. It worked out so well that our
first paper actually became Alex's inaugural paper in the Proceedings of
the National Academy of Sciences when he was elected to the Academy
(Kech\-ris et al., \citeyear{2006}). This work taught me something
about the limitations of fundamental experimental information. We did
some statistical analyses and introduced some funny methods for finding
critically important sites in proteins. Since the crystallographic
structure of the proteins we were studying was known, we looked to see
if the sites we identified statistically were visibly critical to the
structure. Not a chance! So in the end we made our case with statistics,
by saying that the percentage of time that mutations in our sites have
serious consequences is larger than expected by chance, they are closer
to some critical structures then if they were randomly selected and so.
But we never were able to say that if you change the amino acid in one
of the positions we identified, everything collapses.

\begin{figure}[b]

\includegraphics{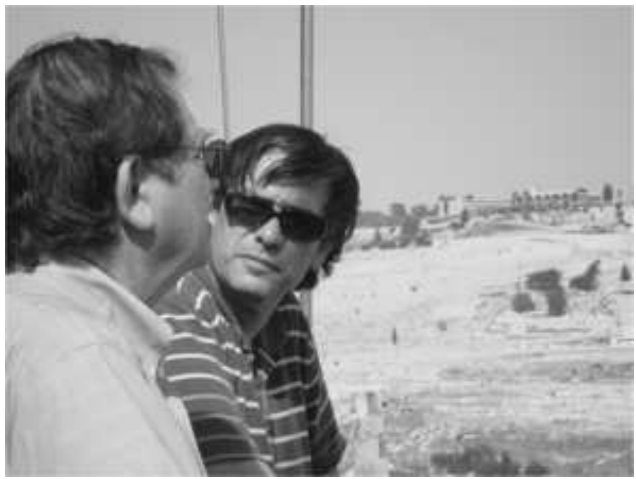}

  \caption{Peter Bickel and Ya'acov Ritov, a day before the interview took place.
  In the background, Mount Oilve, Jerusalem.}
\end{figure}

Then I put together a proposal with Katerina Kech\-ris and a young
colleague, Haiyan Huang, for a special program of the National Science
Foundation, funded largely by the National Institute of General Medical
Sciences, for problems on the borders of biology and mathematics. The
proposal had two parts. One was on a key question of Alex Glazer about
lateral gene transfer between different species of bacteria. A second
part was on the functional importance of genomic sites conserved between
very distant species. We proposed to use a data set from Eddy Rubin's
lab. The reviewers liked the conserved sites part of the proposal, but
hated the lateral gene transfer part, so our overall score was
borderline. Shula Gross was then an NSF program director. She persuaded
the NIGMS people to fund the proposal. Katerina, Haiyan, a student, Na
Xu, and I started working on the conserved sites problem---with very
modest success. Because we got the funding, I~was able to attract a
student, Ben Brown, from an engineering program and put him to work on
the genomics questions. He is both passionate and scholarly about the
biology and has mastered a great range of computing techniques on his
own. Our results on the Rubin data were still not very satisfying.
Fortunately, we were led to change our focus by connecting with a group
at the National Human Genome Research Institute.

Because I have grandchildren in Washington, I decided to find a suitable
academic base for visiting in DC. I looked for a place to work at the
National Human Genome Research Institute. I didn't know anybody there,
but I was on a committee with biochemist Maynard Olson, who had been a
post doc with Alex Glaser. Olson's own post doc, Eric Green, directed a
lab group at NHGRI. There, I got involved in the ENCODE (Encyclopedia of
DNA) project. Little happened during my first visit, but the next year I
brought Ben. That made a big difference. Ben was able to talk with the
biologists in their language and carry through computations.

I had not expected important statistical methods to come out of the
project. My main motivation was learning more about the biology. But, it
turned out that we needed to develop a nonparametric model for the
genome, which might turn out to be interesting and important. This model
is the most nonparametric model you could think of for the genome. In
addition, the method of inference that we developed, a modified block
bootstrap, gives a check on whether what you are doing is reasonable. On
the scales that ENCODE was studying, our theory requires that the
statistics on which our methods are based should have approximately
Gaussian distribution. By plotting the bootstrap distributions of these
statistics, we can see if this assumption is roughly valid. Moreover,
the model is robust in the sense that, even if the approximation is
poor, $p$-values for tests of association between features are
conservative. This work turned out to be very nice theoretically and the
biologists seem to like it a lot. I am now somewhat confident that this
framework may lead to major contributions.

The other direction I've been following is connected with the location
of my second set of grandchildren in Boulder, Colorado. I've been
visiting at the National Center for Atmospheric Research in Boulder,
working with the statistics group headed by Doug Nychka. A basic goal at
NCAR is to do relatively short term weather prediction based on computer
models. You can say ``Why do I need a~computer model? I can make
predictions using yesterday's weather or other past information.'' But,
because of the high dimensionality of the problem, this approach doesn't
work. The computer models are valid enough to dramatically improve
prediction. However, the computer models themselves produce high
dimensional data.

Again, through chance, Thomas Bengtsson came~to Berkeley. Bengtsson had
spent some time in NCAR and, working with the physicists, had been
trying to understand how to use these computer models effectively. They
had hit a serious problem, the collapse of particle filters in high
dimension. Bengtsson, Snyder, Anderson, two physicists at NCAR and I
were able to analyze this phenomenon. This led to a paper in the
\textit{Monthly Weather Review} (Snyder et al.,
\citeyear{2008Snyder}) and some theoretical papers (Bickel, Li and Bengtsson, \citeyear{2008Bickel}). I am
now working with Jing Lei, a graduate student, trying to bypass these
difficulties of particle filters.

As it turns out, both of my current fields of application, genomics and
weather prediction, have fed naturally into my theoretical interests in
understanding high dimensional data analysis. I still work at a~rather
abstract level, and don't deal well with details, but, fortunately, my
students, post docs and colleagues compensate for my shortcomings, so
together we're able to make satisfying contributions to both theory and
practice.

\textit{I want to go back to your student years in Berkeley.}

I started at Caltech, was there for two years, but then transferred to
the University of California, Ber\-keley. I finished my undergraduate work
at Berkeley in one year, because I had done five years of high school in
Canada, not four as in the US. Caltech paid no attention to the extra
year, but Berkeley did give me credit for my fifth high school year,
provided I completed my undergraduate degree in mathematics. I had moved
to UCB intending to switch to psychology. That's what brought me to a
class taught by Joe Hodges. I thought statistics would be necessary for
a psychology student. That class drew
me into statistics just at the time that a graduate class in
mathematical learning theory made me skeptical about psychology.

I took a Master's degree in math while I was deciding whether to go into
math or statistics. I actually wanted to do my Ph.D. with Hodges, but
Hodges insisted that people come to him with their own problems, and I~wasn't ready to do that.
So he steered me to Lehmann. That was very
fortunate for me. Erich Lehmann was really a life guide, not just an
academic one. Academically, my progress was a bit funny. I spent only
two years in my Ph.D. program. I had already taken the basic graduate
probability course. I found statistics interesting and wanted to pursue
it in depth. I could really have gained by studying a little bit more,
but again chance intervened. The Department had an oral Qualifying Exam
for the Ph.D., with three panels of faculty members---in theoretical
statistics, applied statistics and probability theory. The students were
examined by each of the three panels. My friend Helen Wittenberg (now
Shanna Swann) needed a~study partner and enlisted me. So I took the
qualifying exam a~year earlier than I otherwise would have
done.\looseness=1

The applied statistics exam was a bit of a farce. How did one prepare
oneself? One read thoroughly Scheffe's book, \textit{The Analysis of
Variance}, a lovely book, but it's really a theory book, with few
examples of analyses of real data. The panel on applied statistics
consisted of Elizabeth Scott, Jerzy Neyman, Evelyn Fix and Henry
Scheffe. They asked me about the book, and that was OK. Then Betty Scott
actually asked me something applied, and I didn't know what to say. They
passed me anyway. I was tired of school and wanted do a thesis right
away. Erich gave me a problem, which I didn't know very much about, but
I succeeded. The Department hired me; so I stayed.

\textit{How would you describe the people in Berkeley at this time?}

It was a very eminent group and there was a lot of collaboration and a
cordial environment. I have enjoyed these aspects of the department very
much from the beginning. There were many joint papers. Between Hodges
and Lehmann, of course, there was a long collaboration. Blackwell and
Hodges had papers; Blackwell and Le Cam had papers. I don't know about
Neyman and Le Cam, but certainly they interacted intensely. Henry
Scheffe and Erich also worked together a lot before my time. Before he
moved to Stanford in the early 50s, Charles Stein worked with Erich.
There were some tensions in the department, but I wasn't aware of them
at that time.\looseness=1

\begin{figure}[b]

\includegraphics{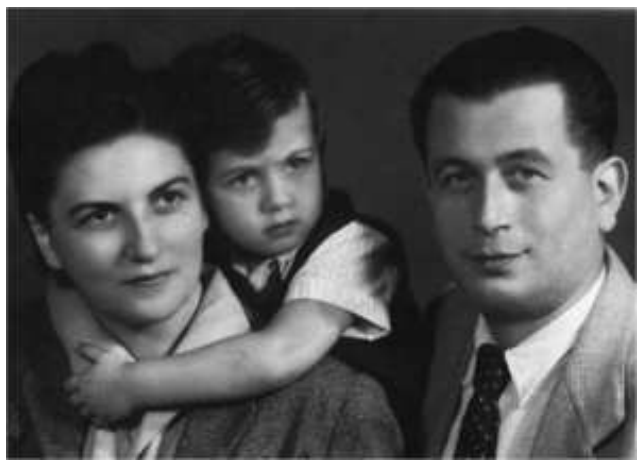}

  \caption{Peter Bickel with his parents, Madeleine and Eliezer Bickel.
Ca. 1943.}
\end{figure}

Le Cam and Neyman viewed themselves as applied statisticians, though the
rest of the statistical world might not have agreed. Betty Scott did
applied statistics, in astronomy and climatology. Hen\-ry Scheffe was a
serious applied statistician. He wor\-ked with Cuthbert Daniel, who was a
private consultant and very impressive. Henry brought him to Berkeley
for a semester of lectures, which was very good for all of us. Joe
Hodges was considered the most talented applied statistician in the
department. He had a wonderful sense of data, but Joe, interestingly
enough, didn't want to be an applied statistician.

The intellectual center of the Department was certainly mathematical
theory. There was a young group of probabilists, including David
Freedman and Lester Dubins and the more senior Loeve and Le Cam. David
Freedman eventually switched to statistics. The relations with Stanford
were excellent. We used to have the Berkeley--Stanford colloquia twice a
quarter, one in Berkeley, and one in Stanford. So, it was a very
pleasant place to work.

I collaborated with Erich and Joe and with David Blackwell. Eventually,
but much later, I collaborated with David Freedman, I think that's about
it with the early years group. Subsequently, I collaborated with later
arrivals, Leo Breiman, Rudy Beran and Warry Millar, as well as, of
course, Kjell Doksum, with whom, in addition to papers, I published
a~book whose second edition we are still working on.

Not long after I started teaching, in the late 60s and early 70s,
Berkeley was full of turmoil. Things happened that had nothing to do
with statistics. I and most of my colleagues supported the Free Speech
Movement. Later on we supported a student strike by holding our classes
off campus, but we were not personally engaged. Of course, it was very
emotional. People left the university from both the right and the left.
When Reagan was governor and when Nixon was president, conflict about
the Vietnam War got heated. I remember teaching class in Dwinelle Hall
at noon, and tear gas coming through the Windows.

\begin{figure}

\includegraphics{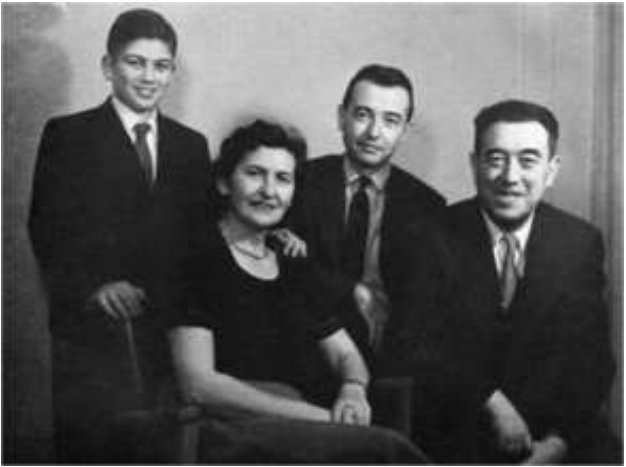}

  \caption{Peter Bickel with his uncle and aunt, Shlomo and Yetta
Bickel, and his cousin Alexander Bickel.}
\end{figure}

\textit{How do you define your generation? Erich Leh\-mann was the leading
person in the second generation. The third generation more or less
started with you and your colleagues.}

To some extent, yes. Moving beyond Berkeley, I've been struck by a
curious observation. A substantial number of leading figures in my
generation came from Caltech. They include, among others, Brad Efron,
Larry Brown, Chuck Stone and Carl Morris. Nobody taught statistics at
Caltech. But, for some reason, we all felt we wanted to do things in the
real world. Among us, only Brad claims that he always wanted to do
statistics. Larry went to Cornell and worked with Jack Kiefer and wrote
a statistical thesis. Chuck said he wanted to do statistics, but he
moved to probability as a student of Karlin. Later he got involved with
Leo Breiman and went into statistics.

\textit{Can you tell about yourself? You once told me, ``We are lucky to
belong to a generation that didn't suffer from wars.'' I found it an
interesting comment from somebody who was born as a Jew in central
Europe during WWII. Can you tell us about your history?}

I was born in Bucharest in 1940, but I was really not very aware of the
war, except that sometimes we had to go to the bomb shelters (when the
Americans were bombing the oil field in Ploesti). Once, when we were
coming back, I saw broken windows in some office buildings. My father
Eliezer (Lothar) Bickel was able to continue to practice medicine during
the war. There was a pogrom in Bucharest, which we narrowly avoided by
my mother Madeleine's courageous behavior. Then, after the war and after
the communists took over, my parents arranged, with difficulty, to leave
Romania legally. But I didn't realize the difficulties at the time.

\begin{figure}

\includegraphics{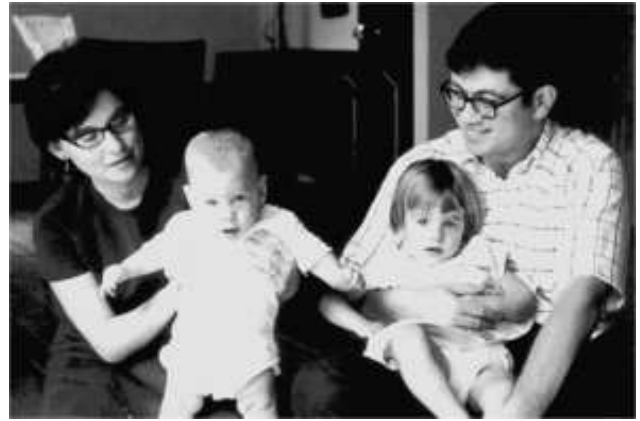}

  \caption{The young Nancy and Peter Bickel, and their daughter and son
Amenda and Stephen. This picture was used as a New Year's greeting card.}
\end{figure}

We went to France in 1948 and then to Canada in 1949. I studied in
France in a public school for ten months. In France my father insisted
on giving me English lessons after an eight hour day of school and
homework. From Canada we went to California. I could have been drafted
in the Vietnam War, but I married young and we had a child. So that
probably was the source of my remark.

\textit{Can you tell us about the intellectual influence of your family
on you?}

My father was born in Bukovina, a German speaking province of the
Austro--Hungarian Empire. He had a traditional Jewish and then a secular
education. In high school, under the influence of one of the high school
teachers, he and other Jewish students became involved in one of the
many intellectual groups of the time. He became, basically, a disciple
of a German philosopher called Constatin Brunner, a son of the grand
rabbi of Hamburg, who rebelled against his father. Brunner was involved
in elaborating a philosophical system based on Spinoza. He believed
strongly that the Jews should assimilate to German culture. One of his
books, called ``\textit{Unser Christus oder Das Wesen des Genie},'' or
in English, ``\textit{Our Christ, or The Essence of Geniu}s,'' among
other things, advocated assimilation to German culture, including
Christianity. Like Brunner, my father favored assimilation. In Romania,
we never celebrated Jewish holidays, and in fact, we celebrated
Christmas, but in a nonreligious way.

My father became the leader of the Brunner group; he was treated almost
as the equivalent of a Hasidic rabbi. He pursued two fields at the same
time---medicine and philosophy. He was able to study medi\-cine at the
University of Bucharest, even though very few Jews were admitted because
there was a~``numerus clausus.'' Also, as he was the second son, his
father wanted him to run the family store and wouldn't pay for his
further education. He rebelled and had to struggle on his own. My father
went to Germany to do post graduate study in medicine. At the same time
he was able to meet and study with Brunner. He was successful in both
fields. I found out later that he was an experimentalist, publishing 23
papers while in Germany. Then, and later, he published books in
philosophy.

If Hitler had not come to power, I might not be here. My father would
have stayed in Germany and become a professor in Berlin. But when Hitler
expelled the foreign Jews, my father returned to Bucha\-rest, married my
mother, and I came to be. My mother was seriously ill during the first
few years of my life, but I had a loving set of grandparents and a nurse
who took care of me and, as far as I can remember, was happy.

I was eleven when my father died. By then we were living in Canada.
Although he was very ill with heart disease, he was studying very hard
to qualify as a doctor in Canada. My relations with him were never easy.
He kept a notebook of anecdotes about me from ages one to five. When I
translated it for my wife Nancy, I saw as I read that the anecdotes are
all instances in which the father humiliates the child. After he died,
I~tried to help my mother, in the house, and by taking a job delivering
for the local drugstore. She coped wonderfully even though her life in
Romania had not prepared her for the role
of relatively poor single mother. Our relations were very close. I would
act as confidante and counselor and was very proud of what help I could
give. She remarried, and that's how we got to California.

I found school work and languages very easy as a~child, too easy, as I
discovered when I got to Caltech and had to compete with many who were
as quick or quicker than I was. I wanted to be a scientist ever since I
read the books of George Gamow. But I was broadly interested in physics
on the one hand and physiology and biochemistry on the other. I avoided
medicine and philosophy, since my father had been a~physician and
philosopher. Fortunately, I found my way through mathematics to
statistics, which has allowed me to dip into almost every
science.

\textit{Your uncle was a lawyer?}

In Romania my uncle Shlomo Bickel was a lawyer, but he, like my father,
and most of his generation of Jews, was part of a movement. He was a
Yiddishist and a Zionist. He was able to get out to the States in 1938.
He couldn't practice law, so became a journalist and wrote weekly
columns for \textit{The Day,} one of the two large Yiddish papers in New
York. He also wrote several books; one chapter of his book ``Rumania,''
later translated into English, was about my father and other rebellious
Bickels. I felt very close to my uncle Shlomo and aunt Yetta. Their
household was full of intellectual and literary discussion. They showed
great affection to each other and to me, particularly after my father
died. They showed me how loving and intellectually lively family life
could be. Like my uncle, I've been fortunate to have a family life full
of love and discussion.

\vspace{2pt}
\section*{Acknowledgments}
\vspace{2pt}

The encouragement of Nancy Bickel was more than helpful both to Peter
and me. The pictures collection was done by her. The pictures were taken
by friends, family, and students and staff in Berkeley and Princeton. I
apologize that I cannot give personal acknowledgments for them.


\end{document}